\documentclass[12pt]{article} 
\newcommand{\Tf}{\Theta_f} 
\newcommand{\F}{{\cal F}} 
\newcommand{\tf}{\theta_f} 
\renewcommand{\u}[1]{\bar{#1}} 
\newcommand{\eqn}[1]{(\ref{#1})} 
\newcommand{\ft}[2]{{\textstyle{\frac{#1}{#2}}}} 
 
\renewcommand{\H}{{\cal H}} 
\begin{document} 
\renewcommand{\theequation}{\thesection.\arabic{equation}} 
\renewcommand{\section}[1]{\addtocounter{section}{1} 
\vspace{7mm} \par \noindent 
  {\bf \sc \thesection . #1}\setcounter{subsection}{0} 
  \par 
  \vspace{2mm} } 
\newcommand{\sectionsub}[1]{\addtocounter{section}{1} 
\vspace{5mm} \par \noindent 
  {\bf \thesection . #1}\setcounter{subsection}{0}\par} 
\renewcommand{\subsection}[1]{\addtocounter{subsection}{1} 
\vspace{2.5mm}\par\noindent {\em \thesubsection . #1}\par 
 \vspace{0.5mm} } 
\renewcommand{\thebibliography}[1]{ {\vspace{5mm}\par \noindent{\bf \sc 
References}\par \vspace{2mm}} 
\list 
 {\arabic{enumi}.}{\settowidth\labelwidth{[#1]}\leftmargin\labelwidth 
 \advance\leftmargin\labelsep\addtolength{\topsep}{-4em} 
 \usecounter{enumi}} 
 \def\newblock{\hskip .11em plus .33em minus .07em} 
 \sloppy\clubpenalty4000\widowpenalty4000 
 \sfcode`\.=1000\relax \setlength{\itemsep}{-0.4em} } 
\font\mybb=msbm10 at 12pt 
\newcommand{\dk}{\delta_\kappa} 
\newcommand{\dkt}{\dk\theta} 
\begin{titlepage} 
\begin{flushright} 
KUL-TF-98/32\\ {\tt hep-th/9809045}\\ \vskip.5cm 
September 1998 
\end{flushright} 
\vspace{.5cm} 
\begin{center} 
\baselineskip=16pt 
\vfill 
{\Large \bf \sc Super M-brane actions in 
adS${}_4$$\,\times\,$S${}^7$ and adS${}_7$$\,\times\,$S${}^4$ } 
\vskip 10.3mm 
{\sc Piet Claus} 
\vskip2cm 
{\it Instituut voor Theoretische Fysica, Katholieke 
Universiteit Leuven,\\ Celestijnenlaan 200D, B-3001 Leuven, Belgium}\\ 
{\tt piet.claus@fys.kuleuven.ac.be} 
\end{center} 
\vfill 
\par 
\begin{center} 
{\sc Abstract} 
\end{center} 
\begin{quote} 
The world-volume action of the M2 brane and the M5 brane in an 
$adS_4\times S^7$ and an $adS_7 \times S^4$ background is derived to 
all orders in anticommuting superspace coordinates $\theta$. 
Contrary to recent constructions of super $p$-brane actions relying on 
supercoset methods, we only use 11 dimensional supergravity torsion and 
curvature constraints. Complete agreement between the two methods is found. 
The possible simplification of the action by choosing a suitable
$\kappa$-gauge is discussed. 
\end{quote} 
\vfill 
\end{titlepage} 
\section{Introduction} 
These days a lot of research is concentrated on the connection between 
string-theory and M-theory on $adS_{p+2} \times S^{D-p-2}$ and 
extended superconformal theories in $p+1$ dimensions \cite{Maldacena, 
GKP,Witten}. 
$adS_5 \times S^5$ describes a maximally supersymmetric vacuum (besides 
flat space) of type IIB-supergravity, which is the near-horizon geometry of 
the D3-brane solution.  For 11 dimensional supergravity we have 
$adS_4\times S^7$ and $adS_7\times S^4$ maximally supersymmetric vacua, 
which are the near-horizon limits of the M2 and M5 solutions resp. 
\par 
The actions of super $p$-brane probes in these near-horizon backgrounds 
are described by a (modified) \lq Dirac'-type term (for $D$-$p$-branes, 
this is the Born-Infeld-type term) and a Wess-Zumino term. 
These actions are invariant under local diffeomorphisms of the world-volume 
and $\kappa$-symmetry and the rigid isometries of the background. 
\par 
One way to realize the superconformal field theory in $p+1$ dimensions 
is to gauge-fix these probe actions in their own near-horizon background 
\cite{CKKTVP}, upon which the isometry group of the background is realized 
as a rigid superconformal symmetry on the world volume.  The bosonic part 
of this project was carried out in \cite{CKKTVP}. 
\par 
Recently the complete super $p$-brane actions in  $adS_{p+2}\times S^{D-p-2}$ 
background geometries have 
been constructed for the superstring \cite{MT,KRR,KR,P,KT} and the D3 brane 
\cite{MT2} in the $adS_5\times S^5$ background solution to 10 dimensional type 
 IIB supergravity  and for the membrane \cite{DFFFTT,dWPPS} in the 
$adS_4 \times S^7$ solution to 11 dimensional supergravity. 
One purpose of this paper is to construct the missing M5-brane action in 
$adS_7\times S^4$, which is an important step in the further understanding 
of the relation between the M5 brane in $adS_7\times 
S^4$ background and the nonlinear six dimensional superconformal 
$(0,2)$ tensor multiplet theory on its world volume \cite{CKVP,Maldacena}. 
\par 
We will consider the near-horizon geometries of the two fundamental branes 
of M-theory.  Firstly the $adS_4\times S^7$ geometry of the M2-brane 
solution, which is given by the metric and the 3-form\footnote{The 
11 dimensional coordinates $X^M$ are split up into horospherical $adS$ 
coordinates $x^{\hat m} = \{x^m,r\}$ and $\xi^{m'}$, which are the coordinates 
on the sphere $S$.} 
\begin{eqnarray} 
ds^2 &=& \left(\frac rR\right)^4 dx^m \eta_{mn} dx^n + \left(\frac 
Rr\right)^2 dr^2 + R^2 d\Omega^2\,,\nonumber\\ 
A_3&=& dx^2 dx^1 dx^0 \left(\frac rR\right)^6\,.\label{metricM2} 
\end{eqnarray} 
Secondly we consider the $adS_7 \times S^4$ near-horizon geometry of the 
M5-brane solution to 11 dimensional supergravity, given by the metric and 
the 3-form 
\begin{eqnarray} 
ds^2 &=& \left( \frac rR \right) d x^m \eta_{mn} dx^n + \left(\frac 
Rr\right)^2 dr^2 + R^2d\Omega^2\,,\nonumber\\ 
A_3 &=& - d\xi^3 d\xi^2 d\xi^1 (3 R^3 \sin^3 \xi_1 \sin^2\xi_2 \sin \xi_3 
\, \xi_4)\,. \label{metricM5} 
\end{eqnarray} 
The solution for $A_3$ can be cast in terms of its magnetic dual $A_6$ 
through the Bianchi identity $dA_6 = \raise4pt\hbox{$\star$}{}d A_3 + 
\ft12 A_3 \wedge d A_3$. For this particular solution the second 
term vanishes and $dA_6 = \raise4pt\hbox{$\star$}{}dA_3$. We can 
choose $A_6$ to be 
\begin{equation} 
A_6 = dx^5\dots dx^0 \left( \frac rR \right)^3\,.\label{formM5} 
\end{equation} 
The covariant field strengths are 
\begin{eqnarray} 
&\mbox{M2}&\qquad F^4_{\u {\hat m}_1\dots \u {\hat m}_4} = - \frac 6R 
\epsilon_{\u {\hat m}_1\dots \u {\hat m}_4}\,,\qquad 
F^7_{\u m'_1\dots m'_7} = \frac 6R \epsilon_{\u m'_1\dots 
m'_7}\,,\nonumber\\ 
&\mbox{M5}&\qquad F^7_{\u {\hat m}_1\dots \u {\hat m}_7} = \frac 3R 
\epsilon_{\u {\hat m}_1\dots \u 
{\hat m}_7}\,,\qquad F^4_{\u m'_1\dots \u m'_4} = \frac 3R \epsilon_{\u 
m'_1\dots \u m'_4}\,. \label{fieldstrengths} 
\end{eqnarray} 
These forms define the lowest order $\theta$ components of the super 4-form 
and 7-form in 11 dimensional superspace. 
\par 
The rest of the paper is organized as follows.  In section~2 we derive the 
vielbeine, connection 1-form and the 4- and 7-forms to all orders in 
$\theta$ for a restricted class of 11 dimensional backgrounds.  This class 
is characterized by the vanishing of the gravitino and covariantly constant 
forms. Contrary to previous work with cosets, we only 
rely on 11 dimensional supergravity torsion and curvature constraints to 
derive these results. We find complete agreement with results obtained from 
the supercoset approach \cite{dWPPS} and provide therefore a nice alternative 
description in supergravity superspace. 
Section~3 gives the actions for the M2 and M5 brane 
in these backgrounds and a detailed proof for the $\kappa$-invariance of 
the M5-brane action is given in section~4. 
In section~5 we discuss the killing-spinor gauge \cite{RKILL} for 
$\kappa$-symmetry and the possible simplification of the action. 
We conclude in section~6 with a short discussion of the results. 
\setcounter{equation}{0} 
\section{A class of 11 dimensional backgrounds} 
In this section we will derive the vielbeine ($E$), spinconnection 
($\Omega$) and forms (${\cal F}_n$), collectively denoted as geometric 
superfields, for a certain class of 11 dimensional backgrounds.  We will 
restrict the background to the case that the gravitino vanishes and the 
basic 11-dimensional superfield ${\cal W}_{\u M_1\dots \u M_4}$ \cite{BH} 
is covariantly constant, which restricts us essentially to the flat 
background (where ${\cal W}$ is vanishing) the $adS_4\times S^7$ and $adS_7 
\times S^4$ background (where ${\cal W}$ is given by \eqn{fieldstrengths}). 
It was shown that these are exact vacua of 11 dimensional supergravity 
\cite{KRa}.  The derivation of the geometric superfields in $adS_4\times 
S^7$ and $adS_7\times S^4$ has also been considered in \cite{dWPPS}, by 
using coset representations.  However, here we will only rely on 
supergravity torsion and curvature constraints. 
\par 
The superspace is parametrized by the coordinates\footnote{We will use 
barred indices $\u M$ to indicate tangent space indices.  Otherwise we 
follow the 11 dimensional conventions given in appendix A.1 of \cite{CKVP}, 
where there was no need to distinguish tangent space indices and curved 
indices as there we worked in flat space.  $\Gamma$-matrices will always 
carry tangent-space indices and we will drop the bars on $\Gamma$-matrix 
indices. For superspace we have the following conventions. 
The components of superforms are defined by $F_n=\ft1{n!} 
dZ^{\Lambda_1}\dots dZ^{\Lambda_n} F_{\Lambda_n\dots \Lambda_1} = \ft1{n!} 
E^{\u \Lambda_1}\dots E^{\u\Lambda_n} F_{\u \Lambda_n\dots \u \Lambda_1}$. 
The essential difference with \cite{CKVP, dWPPS} is 
that the derivative acts from the right, i.e.  $d F_n= dZ^{\Lambda_1}\dots
dZ^{\Lambda_n} dZ^{\Sigma} \partial_\Sigma F_{\Lambda_n\dots \Lambda_1}$ and  
$d (FG) = F dG + (-)^n dF G$ where $G$ is an $n$-form.}
\begin{equation} 
Z^\Lambda= \{ X^M,\theta^A\}\,, 
\end{equation} 
where $X^M$ ($M=0,\dots,10$) are the bosonic coordinates of the 
11 dimensional space and $\theta^A$ ($A=1,\dots,32$) are 32 anticommuting 
coordinates.  The vielbeine and connection satisfy `on shell' torsion and 
curvature constraints \cite{BH} 
\begin{eqnarray} 
{\cal T}^{\u M} &\equiv& d E^{\u M} - E^{\u N} \Omega_{\u N}{}^{\u M} 
= - \bar E \Gamma^{\u M} E \,, \nonumber\\ 
{\cal T}^{\u A} &\equiv& d E^{\u A} - \ft14 (\Omega^{\u M\u N} \Gamma_{MN} 
E)^{\u A}\nonumber\\ 
&=& E^{\u M} (T_{\u M}{}^{\u N_1\dots N_4} F_{\u N_4\dots \u N_1} E)^{\u 
A}\,, 
\label{torsions} 
\end{eqnarray} 
\begin{eqnarray} 
{\cal R}^{\u M\u N} &\equiv& d\Omega^{\u M\u N} - 
\Omega^{\u M}{}_{\u P} \wedge \Omega^{\u P\u N}\nonumber\\ 
&=& \ft12 E^{\u Q} E^{\u P} R_{\u P\u Q}{}^{\u M\u N} + \ft12 \bar E S^{\u M\u 
N\u P_1\dots \u P_4} F_{\u P_4\dots \u P_1} E\,, 
\end{eqnarray} 
where 
\begin{eqnarray} 
T_{\u M}{}^{\u N_1\dots\u N_4} &=& \ft1{288}\left(\Gamma_M{}^{N_1\dots N_4} 
- 8 \delta_M{}^{[N_1} \Gamma^{N_2\dots N_4]}\right)\,,\nonumber\\ 
S_{\u M\u N}{}^{\u P_1\dots P_4} &=& \ft 1{72} \left(\Gamma_{MN}{}^{P_1\dots 
P_4} + 24 \delta_M{}^{[P_1} \delta_N{}^{P_2} \Gamma^{P_3 P_4]}\right)\, 
\end{eqnarray} 
and $R^{\u M\u N}$ and $F$ are the bosonic curvature and 4-form of the 
background. 
\par 
The super 4-form and 7-form are given by 
\begin{eqnarray} 
&&\F_4 = \ft1{4!} E^{\u M_1} \dots E^{\u M_4} F^4_{\u M_4\dots \u M_1} + 
\ft12 E^{\u M_1} E^{\u M_2} \bar E\Gamma_{M_2M_1}E = d{\cal A}_3 
\nonumber\\ 
&& \F_7 = \ft1{7!}E^{\u M_1} \dots E^{\u M_7} F^7_{\u M_7\dots \u M_1} + 
\ft 1{5!}E^{\u M_1} \dots  E^{\u M_5} \bar E \Gamma_{M_5 
\dots M_1}E\,\nonumber\\ 
&&\hspace{.6cm} = d{\cal A}_6 - \ft 12 {\cal A}_3 \wedge \F_4 \label{defn} 
\end{eqnarray} 
The consistency conditions (Bianchi-identities) for the definitions of 
${\cal A}_3$ and ${\cal A}_6$ 
\begin{eqnarray} 0 &=& d \F_4\nonumber\,,\\ 
0 &=& 2 d \F_7 + \F_4 \wedge \F_4\label{bianchidual}\,, 
\end{eqnarray} 
can be checked explicitely using equations \eqn{torsions} and the 
11 dimensional Fierz identities \cite{Fierz}. These conditions also 
determine the relative factors in and between $\F_4$ and $\F_7$. 
\par 
Using coset-space techniques (see e.g.~\cite{cosets}), various backgrounds 
for super $p$-brane actions have been constructed to all orders in $\theta$, 
especially the $adS_4\times S^7$ and $adS_7\times S^4$ near-horizon 
backgrounds in \cite{dWPPS}.  It was established 
that a very powerfull trick to derive the vielbeine and spinconnection 
1-form was to rescale the $\theta$'s, with a parameter that is put to unity 
in the end (see e.g.~\cite{MT,KRR}).  However, this technique is not a 
privilege of coset representations and the same results can be obtained 
using supergravity constraints only.\footnote{In fact for the backgrounds we 
consider, the Maurer-Cartan equations in the coset approach are the 
curvature and torsion constraints in supergravity and they contain enough 
information to derive the geometric superfields to all orders in 
$\theta$.} 
\par 
Consider the transformation 
\begin{equation} 
X^M \rightarrow X^M\,,\qquad \theta^A\rightarrow t \theta^A\,. 
\end{equation} 
Taking the derivative with respect to $t$ of $E$ and $\Omega$ leads to the 
coupled first-order equations in $t$ 
\begin{eqnarray} 
\frac d{dt} E^{\u \Lambda} &=& d (\theta^A E_{A}{}^{\u \Lambda}) - 
\theta^A E_{A}{}^{\u \Delta} E^{\u \Sigma} {\cal T}_{\u \Sigma\u 
\Delta}{}^{\u \Lambda} + E^{\u \Sigma} 
\theta^A \Omega_{A;\u \Sigma}{}^{\u \Lambda} - \theta^A E_{A}{}^{\u \Sigma} 
\Omega_{\u\Sigma} {}^{\u\Lambda}\,,\nonumber\\ 
\frac d{dt} \Omega^{\u M\u N} &=& d(\theta^A \Omega_{A}{}^{\u M\u N}) - \theta^A E_A{}^{\u\Sigma} E^{\u 
\Lambda} {\cal R}_{\u \Lambda \u\Sigma}{}^{\u M\u N} + \Omega^{\u M\u 
P}\theta^A\Omega_{A;\u P}{}^{\u N} - \theta^A \Omega_A{}^{\u M}{}_{\u P} 
\Omega^{\u P\u N}\,.\nonumber\\  \label{dert} 
\end{eqnarray} 
To solve these equations we make the assumption 
\begin{equation} \theta^A E_A{}^{\u M} = \theta^A \Omega_A{}^{\u M\u N} = 
0\, \label{anszats} 
\end{equation} 
and define 
\begin{equation} \Tf^{\u A} = \theta^A E_A{}^{\u A}\,. 
\end{equation} 
Thus \eqn{dert} becomes\footnote{On $E^{\u A}$ we will drop the index most 
of the time.} 
\begin{eqnarray} 
\frac {d}{d t} E &=& (d + \ft14 \Omega^{\u M\u N} \Gamma_{MN} + E^{\u M} 
T_{\u M}{}^{\u N_1\dots \u N_4} F_{\u N_4\dots \u N_1})\Tf\,,\nonumber\\ 
\frac {d}{dt} E^{\u M} &=& 2 \bar \Tf \Gamma^{\u M} E\,,\nonumber\\ 
\frac {d}{dt} \Omega^{\u M\u N} &=& - \bar \Tf S^{\u M\u N \u P_1\dots \u 
P_4} F_{\u P_4\dots \u P_1} E\,.\label{tsystem} 
\end{eqnarray} 
These equations agree completely with the ones derived using coset-space 
techniques \cite{dWPPS} and therefore the two approaches are completely 
equivalent to all orders in $\theta$. 
\par 
Following \cite{dWPPS}, \eqn{tsystem} can be solved straightforwardly, 
by taking multiple derivatives w.r.t.~$t$ and considering the initial 
conditions \cite{KR} 
\begin{eqnarray} 
E|_{t=0} &=& 0\,,\nonumber\\ 
E^{\u M}|_{t=0} &=& e^{\u M} = d X^M e_M{}^{\u M}(X)\,,\nonumber\\ 
\Omega^{\u M\u N}|_{t=0} &=& \omega^{\u M\u N}= d X^M \omega_M{}^{\u M\u 
N}(X)\,, 
\end{eqnarray} 
where $e_M{}^{\u M}$ and $\omega_M{}^{\u M\u N}$ are the vielbein and 
spinconnection components of the bosonic background. 
\par 
The explicit solution reads \cite{dWPPS} 
\begin{eqnarray} 
E &=& \sum_{n=0}^{16} \frac 1{(2n + 1)!} {\cal M}^n 
D\tf\,,\nonumber\\ 
E^{\u M} &=& d X^M e_M{}^{\u M} + 2 \sum_{n=0}^{15} 
\frac 1{(2n + 2)!} \bar \tf \Gamma^{\u M} {\cal M}^n D \tf\,,\nonumber\\ 
\Omega^{\u M\u N} &=& d X^M \omega_M{}^{\u M\u N} - \sum_{n=0}^{15} \frac 
1{(2n + 2)!}\bar \tf S^{\u M\u N\u P_1\dots \u P_4} F_{\u P_4\dots \u P_1} 
{\cal M}^n D\tf\,, 
\end{eqnarray} 
where 
\begin{eqnarray} 
({\cal M})_{\u A}{}^{\u B} &=& 2 (T_{\u M}{}^{\u N_1\dots \u N_4} F_{\u 
N_4\dots \u N_1}\tf)_{\u A} (\bar \tf \Gamma^{M})^{\u B}\nonumber\\ 
&& -\frac14 (\Gamma_{MN} \tf)_{\u A} (\bar \tf S^{\u M\u N\u P_1\dots\u 
P_4} F_{\u P_4\dots \u P_1})^{\u B}\,, 
\label{matrixM} 
\end{eqnarray} 
\begin{equation} 
\tf^{\u A} = \theta^A E_A{}^{\u A}|_{t=0} \equiv \theta^A e_A{}^{\u A}(X)\, 
\end{equation} 
and 
\begin{equation} 
D \tf = \left( d + \ft14 \omega\cdot \Gamma + e^{\u M} T_{\u M}{}^{\u 
P_1\dots \u P_4} F_{\u P_4\dots \u P_1}\right) \tf\,.\label{kill} 
\end{equation} 
It is straightforward to verify that \eqn{anszats} is fulfilled, using 
symmetry properties of 11 dimensional $\Gamma$ matrices, and therefore 
the derivation is consistent. 
\par 
As a side remark we note that part of \eqn{kill} is the killing spinor 
equation 
\begin{equation} 
0=\delta \psi_M = (\partial_M + \ft14 \omega_M \cdot \Gamma + e_M{}^{\u M} 
T_{\u M}{}^{\u P_1\dots \u P_4} F_{\u P_4\dots \u P_1})\epsilon\,. 
\end{equation} 
The solution to these equations can be written as 
\begin{equation} 
\epsilon_{kill}^{\u A}(X) = \epsilon_0^A {\cal K}_A{}^{\u 
A}(X)\,,\label{defK} 
\end{equation} 
with $\epsilon_0^A$ a constant 11 dimensional spinor. 
Therefore we can simplify the expressions for the vielbeine and connection 
1-form by taking \cite{KRR} 
\begin{equation} 
e_A{}^{\u A} = {\cal K}_A{}^{\u A}\,\label{killspin} 
\end{equation} 
and it follows that 
\begin{equation} 
D\tf^{\u A} \rightarrow (d\theta^A) {\cal K}_A{}^{\u A}\,. 
\end{equation} 
This gauge was for obvious reasons called the killing-spinor gauge 
and should be thought of as an alternative for the Wess-Zumino 
gauge $e_A{}^{\u A}= \delta_A{}^{\u A}$.  In fact, for flat space it 
reduces to the Wess-Zumino gauge, because the flat-space killing spinors 
are constant spinors. In this killing-spinor gauge we have that
\begin{equation}
E_M{}^{\u M} (Z)= e_M{}^{\u M}(X)\,,\qquad E_M{}^{\u A}(Z) = 0\,,
\end{equation}
which means that there are no higher order $\theta$ corrections to the
bosonic solution for the vielbein $e_M{}^{\u M}(X)$ and the gravitino $\Psi_M(X)$.
\par 
We can also obtain expressions for ${\cal A}_3$ and ${\cal A}_6$ to all 
orders in $\theta$, by considering $t$-derivatives of the 
forms. Using \eqn{tsystem} and Fierz-identities it is non-trivial to derive 
that 
\begin{equation} 
\frac{d}{dt} \F_4 = d (E^{\u M_1} E^{\u M_2} 
\bar E \Gamma_{M_2M_1} \Theta_f)\,, 
\end{equation} 
which we can easily integrate to find 
\begin{equation} 
{\cal A}_3 = \frac 1{3!} e^{\u M_1} e^{\u M_2} e^{\u M_3} A_{\u M_3\u M_2\u M_1} + 
\int_0^1 dt (E^{\u M_1} E^{\u M_2} \bar E \Gamma_{M_2M_1} 
\Tf)\,.\label{3form} 
\end{equation} 
In the same way, one finds that 
\begin{equation} 
\frac{d}{dt} \F_7 = \frac 2{5!} d\left(E^{\u M_1}\dots E^{\u M_5} 
\bar E \Gamma_{M_5\dots M_1} \Tf \right) - E^{\u M_1} E^{\u M_2} \bar E 
\Gamma_{M_2M_1} \Tf \wedge \F_4\,, 
\end{equation} 
which can be integrated to 
\begin{eqnarray} 
{\cal A}_6\hspace{-2mm} &=&\hspace{-2mm} \frac{1}{6!}e^{\u M_1} \dots e^{\u M_6} 
A_{\u M_6 \dots \u M_1} +\nonumber\\ 
\hspace{-2mm}&&\hspace{-2mm}\int_0^1 dt \left(\frac 2{5!} E^{\u M_1} \dots E^{\u M_5} 
\bar E \Gamma_{M_5\dots M_1} \Tf + \frac12 {\cal A}_{3} \wedge 
E^{\u M_1} E^{\u M_2} \bar E \Gamma_{M_2M_1} \Tf\right)\,,\label{6form} 
\end{eqnarray} 
where under the $t$-integrals we mean the rescaled vielbeine and $3$-form. 
\par 
In flat space the matrix ${\cal M}$ vanishes because the field strengths 
vanish and therefore in the Wess-Zumino gauge the vielbeine are given by 
\begin{equation} 
E = d\theta\,,\qquad E^{M} = dX^M + \bar \theta \Gamma^M 
d\theta\,.\label{flatviel} 
\end{equation} 
It is reassuring to see that using these results in the equations for 
${\cal A}_3$ and ${\cal A}_6$, the flat forms (see e.g. \cite{CKVP}) are 
obtained \cite{dWPPS}. 
\setcounter{equation}{0} 
\section{The M-brane actions} 
Super $p$-brane actions consist in general of two parts 
\begin{equation} 
I = I_{kin} + I_{WZ}\,. 
\end{equation} 
$I_{kin}$ contains the induced metric 
\begin{equation} 
g_{\mu\nu} = E^{\u M}_\mu \eta_{MN} E^{\u N}_\nu\,; \qquad g\equiv -\det 
g_{\mu\nu}\,, 
\end{equation} 
and is an integral over the world volume (with coordinates 
$\sigma^\mu$). 
The superspace coordinates $Z^\Lambda = Z^\Lambda(x)$ are taken to be 
fields on the world volume and by $E_\mu^{\u M}$ we mean the pull back 
to the world volume of $E^{\u M}$.  The Wess-Zumino component is an 
integral of a closed $p+2$-form over a $p+2$ dimensional manifold that has 
the world volume as its boundary.  
\par 
To be more specific for M2 we have (the two signs correspond to branes and
anti-branes)
\begin{equation} 
I_{kin} = - \int_{{\cal M}_3} d^3\sigma \sqrt g\,, \qquad I_{WZ} = 
\mp \int_{{\cal M}_4} \F_4\,, 
\end{equation} 
where $\F_4$ is the closed background super 4-form. 
\par 
This action is invariant under local diffeomorphisms and $\kappa$-symmetry 
and by construction under the rigid isometries of the background, 
especially $OSp(8|4)$ for $adS_4\times S^7$. After gauge-fixing the local 
symmetries of this action we will end up with a superconformal field theory 
in 3 dimensions as $OSp(8|4)$ is the superconformal group in 3 dimensions. 
\par 
The M5 world-volume action is somewhat more involved. We will use the 
covariant formulation of \cite{5b}, following 
\cite{CKVP}, where the derivation of the M5-brane action 
in flat 11 dimensional superspace was given in detail. 
The covariance is achieved by introducing an extra real scalar 
degree of freedom, which is pure gauge for some additional gauge 
transformation.  Besides the superspace coordinates, we also introduce an 
antisymmetric tensor $B_{\mu\nu}(\sigma)$ and a real scalar $a(\sigma)$, 
which live only on the world volume. The two components of the action are 
\begin{eqnarray} 
I_{kin} &=& \int_{{\cal M}_6} d^6 \sigma \left(-\sqrt{-\det(g_{\mu\nu} + i 
\H^*_{\mu\nu})} 
\mp \frac{\sqrt g}{4} \H^{*\mu\nu} \H_{\mu\nu}\right)\nonumber\\ 
&\equiv& -\int_{{\cal M}_6} 
d^6 \sigma \left({\cal G} \pm \frac{\sqrt g}{4} \H^{*\mu\nu} 
\H_{\mu\nu}\right) 
\end{eqnarray} 
and 
\begin{equation} 
I_{WZ} = \mp \int_{{\cal M}_7} I_7\,. 
\end{equation} 
We defined the following fields\footnote{We took some different definitions 
w.r.t.~\cite{CKVP}.} 
\begin{eqnarray} 
&& H=dB\,,\qquad {\cal H}=H + {\cal A}_3\,, \nonumber\\ 
&& u_\mu=\partial_\mu a\,,\qquad u^2=u_\mu g^{\mu\nu}u_\nu\,,\qquad 
v_\mu=\frac{u_\mu}{\sqrt{u^2}}\,,\nonumber \\ 
&& {\cal H}_{\mu\nu}=v^\rho{\cal H}_{\mu\nu\rho}\,, \qquad 
{\cal H}^*_{\mu\nu}=v^\rho{\cal H}^*_{\mu\nu\rho}\,,\nonumber\\ 
&&{\cal H}^*_{\mu\nu\rho}=\frac{\sqrt{g}}{6} 
\epsilon_{\mu\nu\rho\sigma\tau\phi} {\cal H}^{\sigma\tau\phi}\,, \nonumber \\ 
&&I_7 = \F_7 + \ft1{2} {\cal H}\wedge \F_4\,. 
\end{eqnarray} 
In all these expressions it is understood that the forms have been pulled 
back to the world volume and indices $\mu$, $\nu$ are raised and lowered 
with the induced metric $g_{\mu\nu}$.  By construction $I_7$ is a total 
derivative 
\begin{equation} 
I_7 = d\left({\cal A}_6 + \ft12 B \wedge \F_4\right) =d\left({\cal A}_6 + 
\ft12 H\wedge {\cal A}_3 \right) 
\end{equation} 
and its pull-back can be integrated over a 7 dimensional space ${\cal M}_7$ 
with the world volume as its boundary. 
\par 
The gauge invariances of the M5-brane action are diffeomorphisms of the world 
volume, $\kappa$-symmetry (discussed in more detail in the next section) 
and the tensor gauge symmetry and PST-gauge symmetries\footnote{The 
normalization of the Wess-Zumino term can be fixed by requiring invariance 
under III \cite{CKVP}.} \cite{5b, CKVP}, I,II and III, which only act on 
$B_{\mu\nu}(x)$ and $a(x)$ 
\begin{eqnarray} 
&{\rm (I)}&\quad \delta_I B_{\mu\nu} = 2 \partial_{[\mu} 
\Lambda_{\nu]}\,,\qquad \delta_I a=0\,;\nonumber\\ 
&{\rm (II)}&\quad \delta_{II} B_{\mu\nu} = \frac 1{\sqrt{u^2}} \varphi 
\left(\H_{\mu\nu} \pm 2 \frac{\delta {\cal G}}{\delta 
\H^{*\mu\nu}}\right)\,,\qquad \delta_{II} a=\varphi\,;\nonumber\\ 
&{\rm (III)}&\quad \delta_{III} B_{\mu\nu} = \psi_{[\mu} v_{\nu]}\,,\qquad 
\delta_{III} a = 0\,. 
\end{eqnarray} 
\par 
By construction, the action is invariant under the rigid superisometries of 
the 11 dimensional background, which is $OSp(6,2|4)$ for $adS_7\times S^4$ 
and super-Poincar\'e transformations in $(1,9)$ dimensions for flat space. 
Because $OSp(6,2|4)$ is the superconformal group in 6 dimensions, after gauge 
fixing local symmetries, the complete non-linear interacting world-volume 
action will be superconformally invariant. For the flat case it was established 
in \cite{CKVP} that only the linearized world-volume theory is 
superconformally invariant. 
\setcounter{equation}{0} 
\section{$\kappa$ symmetry of the M5-brane action} 
In this section we will proof that the M5-brane action constructed in the 
previous section is $\kappa$-symmetric.  The $\kappa$-symmetry has been 
proven in flat space both in the non-covariant form \cite{APPS} and in 
covariant form \cite{5b} (see also \cite{CKVP} for a detailed proof). 
Unlike for $D$-$p$ branes \cite{BT} and membranes \cite{BST}, no complete 
detailed proof for the $\kappa$-invariance of the M5 brane in a generic 
11 dimensional supergravity background, has been given.  In this section we 
will provide such a proof for a background with vanishing gravitino and 
covariantly constant forms, based on exactly the same strategy as in the 
flat background (see \cite{CKVP}, appendix C). 
\par 
Given the variations $\delta Z^\Lambda$ of the world-volume fields 
$Z^\Lambda = \{X^M, \theta^A\}$, we define\footnote{From a computational 
point of view this derivation is essentially the same as in section 2.} 
\begin{eqnarray} 
\delta \widehat E^{\u \Lambda} &\equiv& \delta Z^\Lambda 
E_\Lambda{}^{\u \Lambda}\nonumber\\ 
\delta \widehat \Omega_{\u \Lambda}{}^{\u \Sigma} &\equiv& \delta Z^\Lambda 
\Omega_{\Lambda;\u \Lambda}{}^{\u \Sigma}\,,\label{defdelta} 
\end{eqnarray} 
which are $0$-forms\footnote{In order not to confuse here with the 
variation of the $1$-forms $\delta (E^{\u\Lambda})$, we use hats when we 
mean the $0$-form.}. A universal feature of $\kappa$-symmetry is 
\begin{equation} 
\dk E^{\u M} =0\,,\label{feature} 
\end{equation} 
which allows one to express $\dk X^M$ in terms of $\dk \theta^A$. 
It follows that 
\begin{equation} 
\delta E^{\u \Lambda} = d(\delta \widehat E^{\u \Lambda}) - \delta \widehat 
E^{\u \Delta} E^{\u \Sigma} {\cal T}_{\u\Sigma\u\Delta}{}^{\u \Lambda} + 
E^{\u\Sigma} \delta \widehat \Omega_{\u\Sigma}{}^{\u\Lambda} - \delta 
\widehat E^{\u \Sigma} \Omega_{\u\Sigma}{}^{\u\Lambda}\,,\label{lemma} 
\end{equation} 
which using \eqn{torsions} and \eqn{feature} reduces to 
\begin{eqnarray} 
\delta E^{\u M} &=& - 2 \bar E \Gamma^M \delta\widehat E + {\rm 
Lorentz}\nonumber\\ 
\delta E^{\u A} &=& d(\delta \widehat E^{\u A}) + E^{\u 
M} (T_{\u M}{}^{\u N_1\dots \u N_4} F_{\u N_4\dots N_1} \delta \widehat 
E)^{\u A} + {\rm Lorentz} 
\end{eqnarray} 
Also we define the pull-backs of the $\Gamma$-matrices 
\begin{equation} 
\gamma_\mu = \Gamma_M E_\mu^{\u M}\,,\quad \gamma_{\mu\nu}= \gamma_{[\mu} 
\gamma_{\nu]}\,, \dots\, . 
\end{equation} 
\par 
To demonstrate the $\kappa$-invariance of the action we begin by 
considering the variation of the different terms in the action.  Since all 
building blocks of the action are manifestly Lorentz invariant we don't 
have to care about the last two terms in \eqn{lemma}.  \footnote{In the 
following expressions we drop the hats again and all $\delta E^{\u 
\Lambda}$ are defined as in \eqn{defdelta}} 
\begin{equation} 
\dk g_{\mu\nu} = 4\dk {\bar E} \gamma_{(\mu} E_{\nu)}\,;\qquad \dk 
\sqrt{g}= 2 \sqrt{g} \dk {\bar E} \gamma^\mu E_\mu\,. 
\end{equation} 
For the 4 and 7 forms, $\F_4$ and $\F_7$, we find using 
\eqn{lemma} and \eqn{torsions} again, that 
\begin{equation} 
\dk \F_4 = d (E^{\u M_1} E^{\u M_2} \bar E \Gamma_{M_2M_1} \dk E)\,. 
\end{equation} 
As $d$ and $\dk$ commute, this means 
\begin{equation} 
\dk {\cal A}_3 = E^{\u M_1} E^{\u M_2} \bar E \Gamma_{M_2M_1} \dk E + d 
{\cal A}_2\, 
\end{equation} 
and therefore we can `define' the $\kappa$-transformation of $B$ to be 
\begin{equation} 
\dk B = - {\cal A}_2\,, 
\end{equation} 
where the pull-back to the world volume of ${\cal A}_2$ is understood. 
The explicit form of ${\cal A}_2$ will of course depend on the choice of 
${\cal A}_3$, but an explicit derivation of ${\cal A}_2$ is beyond the 
scope of this paper.  It follows that 
\begin{equation} 
\dk \H = E^{\u M_1} E^{\u M_2} \bar E \Gamma_{M_2M_1} \dk E\,. 
\end{equation} 
Again using \eqn{torsions} and \eqn{lemma} one can derive that 
\begin{equation} 
\dk \F_7 = \frac 2{5!} d\left(E^{\u M_1}\dots E^{\u M_5} 
\bar E \Gamma_{M_5\dots M_1} \dk E \right) - E^{\u M_1} E^{\u M_2} \bar E 
\Gamma_{M_2M_1} \dk E \wedge \F_4\, 
\end{equation} 
and therefore 
\begin{equation} 
\dk I_7 = d\left(\frac 2{5!} E^{\u M_1}\dots E^{\u M_5} 
\bar E \Gamma_{M_5\dots M_1} \dk E + \frac12 \H \wedge E^{\u M_1} E^{\u 
M_2} \bar E \Gamma_{M_2M_1} \dk E \right)\,. 
\end{equation} 
For the PST scalar $a$ we make the ansatz that $\dk a=0$. 
\par 
Having established the $\kappa$-symmetry of the basic components in the action, 
the rest of the proof is exactly the same as in flat space (appendix C of 
\cite{CKVP}).  We will repeat some of the details here to make this proof 
self-contained. 
\begin{eqnarray} 
\dk \sqrt{u^2}&=&-\ft12\sqrt{u^2} v^\mu v^\nu \dk g_{\mu\nu}\,,\qquad 
\dk v_\mu=\ft12 v_\mu v^\nu v^\rho \dk g_{\nu\rho}\nonumber\\ 
\dk{\cal H}^{*\mu\nu}&=&\frac{1}{\sqrt{g}} 
\epsilon^{\mu\nu\rho\sigma\tau\phi} v_\rho \bar E_\sigma 
\gamma_{\tau\phi}\dk E - {\cal H}^{*\mu\nu} \left( 2 \bar E_\rho 
\gamma^\rho \dk E +\ft12 v^\rho v^\sigma\dk g_{\rho\sigma}\right)\,, 
\nonumber\\ 
\dk {\cal H}_{\mu\nu}&=&\left( \dk {\cal H}_{\mu\nu\rho}\right) 
v^\rho - {\cal H}_{\mu\nu}\frac{ \dk \sqrt{u^2}}{ \sqrt{u^2}}+ {\cal 
H}_{\mu\nu\rho}v_\sigma \dk g^{\rho\sigma}\,. 
\end{eqnarray} 
Now we define 
\begin{eqnarray} 
\dk\left( \mp \int d^6 x \frac{\sqrt{g}}4 {\cal H}^{*\mu\nu} {\cal H}_{\mu\nu} 
+I_{WZ} \right) &=& \mp \frac{\sqrt{g}}{2}\bar E_\mu T^\mu\dk E\,, 
\nonumber\\ 
\dk{\cal G}&=&-\frac{g}{2{\cal G}} \bar E_\mu U^\mu\dk E\, 
\end{eqnarray} 
and 
\begin{equation} 
\bar \gamma = \frac1{6!\sqrt g} 
\epsilon^{\mu_1\dots\mu_6} \gamma_{\mu_1\dots\mu_6} \quad 
\Rightarrow \quad \bar \gamma^2=1\,. 
\end{equation} 
The tensors $T^\mu$ and $U^\mu$ are exactly the same as in the flat case 
and their explicit expressions can be found in equations (C.6) and (C.9) of 
\cite{CKVP}.  It was established that the $T^\mu$ and $U^\mu$ are related 
by 
\begin{equation} 
U^\mu = T^\mu \rho \ {\rm where} \ \rho = \bar \gamma + 
\ft{1}2{\cal H}_{\mu\nu}^*v_\rho \gamma^{\mu\nu\rho} - 
\ft1{16\sqrt{g}}\epsilon^{\mu\nu\rho\sigma\tau\phi} {\cal H}^*_{\mu\nu} 
{\cal H}^*_{\rho\sigma} \gamma_{\tau\phi}\, 
\end{equation} 
and 
\begin{equation} 
\rho^2=\frac{{\cal G}^2}{g}\,. \label{rho2} 
\end{equation} 
Defining 
\begin{equation} 
\Gamma^{({\rm M5})} = \frac{\sqrt g}{{\cal G}} \rho, 
\end{equation} 
the action is $\kappa$-invariant if 
\begin{equation} 
\dk E = (1\pm\Gamma^{({\rm M5})})\kappa(\sigma)\,, 
\qquad \dk E^{\u M}=0\,, 
\end{equation} 
with $\kappa$ a generic 11 dimensional $\sigma$-dependent spinor. 
We repeat the expression for $\Gamma^{({\rm M5})}$ 
\begin{equation} 
\Gamma^{({\rm M5})} = \frac{1}{{\cal G}} \left( \sqrt{g} \bar \gamma + 
\ft{\sqrt{g}}2{\cal H}_{\mu\nu}^*v_\rho \gamma^{\mu\nu\rho} - 
\ft1{16}\epsilon^{\mu\nu\rho\sigma\tau\phi} {\cal H}^*_{\mu\nu} {\cal 
H}^*_{\rho\sigma} \gamma_{\tau\phi}\right)\,.  \end{equation} 
It satisfies $\Gamma^2=1$ and ${\rm tr}\, \Gamma=0$.  Therefore 
we have established the $\kappa$-invariance of the M5-brane action 
action in a 11 dimensional background with vanishing gravitino and 
covariantly constant forms. 
\par 
For completeness and future reference we also give the 
$\kappa$-transformation in the M2 case \cite{BST}. 
This action is invariant under 
\begin{equation} 
\dk E^{\u A} = (1 \pm \Gamma^{({\rm M2})})\kappa(\sigma)\,, 
\qquad \dk E^{\u M} = 0\,, 
\end{equation} 
where 
\begin{equation} 
\Gamma^{({\rm M2})} = \frac{\varepsilon^{\mu\nu\rho}}{3!\sqrt g} 
\gamma_{\mu\nu\rho}\,. 
\end{equation} 
\setcounter{equation}{0} 
\section{Simplifying the action by $\kappa$-gaugefixing} 
In the previous section we derived the action to all orders in $\theta$. 
For the near-horizon backgrounds, the vielbeine contain terms up to 
order 32 in $\theta$ and the explicit expression for the action is 
therefore very complicated.  As discussed in \cite{RKILL}, we expect that a
suitable gauge-fixing of the $\kappa$-symmetry related to the killing
spinors will simplify the action dramatically.  It is the purpose of this
section to show by explicit computation that this is indeed the case  
\par 
First of all we need an expression for the matrix ${\cal K}$ in \eqn{defK}. 
This can be obtained by explicitely solving the killing-spinor equations. 
For $adS_4\times S^7$ (M2 source),  using the coordinates defined in 
\eqn{metricM2}, the killing-spinor equations take the form 
\begin{eqnarray} 
0&=&\delta \psi_m = \partial_m\epsilon +  \left(\frac 
rR\right)^2 \frac 1R \Gamma_m\Gamma_r (1 - \Gamma^{012})\epsilon\,,\nonumber\\ 
0&=&\delta \psi_r = \partial_r \epsilon - \frac 1r \Gamma^{012}\epsilon\,, 
\nonumber\\ 
0&=&\delta \psi_{m'} = \tilde \nabla_{m'}\epsilon  - \frac 1{2} 
\Gamma_r \tilde e_{m'}{}^{\u m'} \Gamma_{m'} \Gamma^{012}\epsilon\,, 
\end{eqnarray} 
where $\tilde \nabla_{m'}$ is the covariant derivative 
and $\tilde e_{m'}{}^{\u m'}$ the vierbein on the unit 7-sphere. 
The first equation suggests to introduce projections 
of a generic 11 dimensional Majorana spinor $\lambda$ 
\begin{equation} 
\lambda_\pm = {\cal P}^{({\rm M2})}_{\pm} \lambda = \ft12(1 \pm \Gamma^{012}) 
\lambda\,.\label{projM2} 
\end{equation} 
Acting with this projector\footnote{The matrix 
$\Gamma^{012}$ has the properties ${}[\Gamma^{012}, \Gamma^m]=0$, 
$\{\Gamma^{012},\Gamma^r\}=0$ and $\{\Gamma^{012},\Gamma^{m'}\} = 0$. 
Also $(\Gamma^{012})^T = {\cal C}^{-1} \Gamma^{012} {\cal C}$, with 
${\cal C}$ the 11 dimensional charge-conjugation matrix.  This means 
that $\bar \lambda_{\pm}= \pm \bar \lambda_{\pm} \Gamma^{012}$ 
with $\bar \lambda = \lambda^T {\cal C}$.} we get 6 killing equations 
\begin{eqnarray} 
0&=&\partial_m \epsilon_-\,, \nonumber\\ 
0&=&\partial_m \epsilon_+ + 2 \left(\frac rR\right)^2 \frac 1R 
\Gamma_{mr} \epsilon_-\,,\nonumber\\ 
0&=&\partial_r \epsilon_\pm \mp \frac1r 
\epsilon_{\pm}\,,\nonumber\\ 
0&=&\tilde \nabla_{m'} \epsilon_\pm \mp \frac1{2} \Gamma_r \tilde 
e_{m'}{}^{\u m'} \Gamma_{m'} \epsilon_{\pm}\,. 
\end{eqnarray} 
It is straightforward to solve these equations\footnote{The solution for 
the killing equation on the sphere can be found in \cite{LPR}} and the explicit 
solution reads 
\begin{eqnarray} 
\epsilon_- &=& \left( \frac Rr\right) f^-_{(\rm M2)}(\xi) 
\epsilon^0_-\,,\nonumber\\ 
\epsilon_+ &=& \left( \frac rR\right) f^+_{(\rm M2)}(\xi)\left[ 
\epsilon^0_+ - 2 \frac {x^m}{R} 
\Gamma_{mr}\epsilon_-^0\right]\,, 
\label{killingM2} 
\end{eqnarray} 
where 
\begin{equation} 
f^\pm_{(\rm M2)}(\xi) = (\cos \ft{\xi_1}2 \pm \Gamma_{r\xi_1} \sin
\ft{\xi_1}2)\prod_{k=2}^7 
(\cos \ft{\xi_k}2 + \Gamma_{\xi_{k-1}\xi_k} \sin \ft{\xi_k}2)\,,\label{fM2} 
\end{equation} 
and $\epsilon_\pm^0$ are the two projections of a constant majorana spinor, 
which make 32 killing spinors, and provides us with the matrix ${\cal K}$ 
defined in \eqn{defK}. 
\par 
In the same way one establishes for $adS_7\times S^4$ (M5 source) that the 
killing spinors can be written as 
\begin{eqnarray} 
\epsilon_- &=& \left( \frac Rr\right)^{1/4} f^-_{(\rm M5)}(\xi) 
\epsilon^0_-\,,\nonumber\\ 
\epsilon_+ &=& \left( \frac rR\right)^{1/4} f^+_{(\rm M5)}(\xi)\left[ 
\epsilon^0_+ - \frac12\frac {x^m}{R}\Gamma_{mr} \epsilon_+^0\right]\,, 
\label{killingM5} 
\end{eqnarray} 
where 
\begin{equation} 
f^\pm_{(\rm M5)}(\xi) = (\cos \ft{\xi_1}2 \pm \Gamma_{r\xi_1} \sin
\ft{\xi_1}2)\prod_{k=2}^4 
(\cos \ft{\xi_k}2 + \Gamma_{\xi_{k-1}\xi_k} \sin \ft{\xi_k}2)\,.\label{fM5} 
\end{equation} 
Now\footnote{The matrix $\Gamma^{012345}$ satisfies $\{\Gamma^{012345}, 
\Gamma^m\}=0$, ${}[\Gamma^{012345},\Gamma^r]=0$ and 
$[\Gamma^{012345},\Gamma^{m'}] = 0$.  Also $(\Gamma^{012345})^T = - {\cal 
C}^{-1} \Gamma^{012345} {\cal C}$.  This means that $\bar \epsilon_{\pm}= 
\mp 
\bar \epsilon_{\pm} \Gamma^{012345}$.} 
\begin{equation} 
\epsilon_\pm= {\cal P}^{({\rm M5})}_\pm \epsilon= \ft12 (1 \pm \Gamma^{012345})\epsilon 
\label{projM5} 
\end{equation} 
defines the projections for M5. 
\par 
Next we will consider the following gauges for $\kappa$-symmetry
\begin{eqnarray}
&(i)&\qquad {\cal P}_- \tf = \tf{}_- =  0\,,\nonumber\\
&(ii)&\qquad {\cal P}_+ \tf = \tf{}_+ =  0\,,
\label{kappagauges}
\end{eqnarray}
where the appropriate projector \eqn{projM2} or \eqn{projM5} has to be taken
for the near-horizon geometry one considers. 
\par
Firstly take case $(i)$, which is the situation considered in \cite{RKILL}.
For the $adS_4 \times S^7$-supergeometry in the killing-spinor gauge
\eqn{killspin} we have
\begin{equation} 
\left.(D \theta_f)\right|_{\tf{}_- = 0} = \left(\frac rR\right)^{1/2w} f^+ 
d\theta_{+}\,, 
\end{equation} 
introducing $w=\ft12$ and $w=2$ for M2 and M5
resp. \cite{CKKTVP}, 
and it follows that ${\cal M} D\tf$ becomes zero in this gauge. Indeed we 
compute 
\begin{eqnarray} 
\left.({\cal M} D\tf)\right|_{\tf{}_-=0} &=& - \frac12 \Gamma_{rm} \Gamma^{012}
\tf{}_{+}\, \bar 
\tf{}_{+} \Gamma^m ({\cal K} d\theta)_{+} F_{\u 0\u 1\u 2\u 
r}\nonumber\\ 
&&+\frac12 \Gamma_{rm} \tf{}_{+}\, \bar \tf{}_{+} \Gamma^m \Gamma^{012} 
({\cal K} d\theta)_{+} F_{\u 0\u 1\u 2\u 
r}\,, 
\end{eqnarray} 
which clearly vanishes.  Also for the $adS_7 \times S^4$-supergeometry
explicit computation shows that this is the case in this gauge. 
Therefore the vielbeine and forms simplify dramatically. 
Defining $\lambda=\theta_+$, we find in both cases that 
\begin{eqnarray} 
E^{\u m} &=& (dx^m  + \bar \lambda \Gamma^m d\lambda) e_{m}{}^{\u 
m} \,,\nonumber\\ 
E^{\u r} &=& dr e_r{}^{\u r}\,,\nonumber\\ 
E^{\u m'} &=& d\xi^{m'} e_{m'}{}^{\u m'}\,,\nonumber\\ 
E &=& (-g_{00})^{\ft14} f^+(\xi) d\lambda\,, 
\end{eqnarray} 
where $e_{\hat m}{}^{\u{\hat m}}$ and $e_{m'}{}^{\u m'}$ are the vielbeine 
of the $adS$-space and the sphere resp. $g_{00}$ is the $00$-component of 
the metric. The expressions for the vielbeine 
are as simple as for the flat background \eqn{flatviel}. 
The forms can then be found by plugging in these vielbeine in 
\eqn{3form}-\eqn{6form} and the action formulas follow. 
\par 
Secondly we can consider case $(ii)$. 
We now apply the killing-spinor gauge 
\eqn{killspin} combined with $\tf{}_+ = 0$.
It follows that 
\begin{equation} 
\left.(D\tf)\right|_{\tf{}_+=0}= \left(\frac Rr \right)^{1/2w} f^-(\xi) 
d\theta_{-} + dx^m \left(\frac rR\right)^{1/w} \frac 1{wR} \Gamma_{mr} 
\tf{}_-\,. 
\end{equation} 
For both the M2 and M5 near-horizon geometries it follows that  
\begin{equation} 
\left.({\cal M}D\tf)\right|_{\tf{}_+=0} = {\cal M} \left(\frac
rR\right)^{1/w} \frac 1{wR} \Gamma_{mr}\tf{}_-\,, \qquad 
\left.({\cal M}^2D\tf)\right|_{\tf{}_+=0} = 0\,, 
\end{equation} 
by explicit computation. In this gauge the vielbeine contain terms up to
order $(\theta_-)^4$.
\par
Whether these gauges are admissible, i.e. if they are compatible with the
reparametrization gauge and the classical solution to the brane-wave
equations one wishes to take, 
has to be considered for each case at hand and goes beyond the
scope of this paper.
\section{Discussion} 
In this paper the super M-brane actions were derived to all orders in 
anticommuting variables $\theta$ for specific backgrounds.  The main 
new result is the complete derivation of the M5-brane action in its 
near-horizon background.  It was shown that the actions can be determined 
completely relying only on supergravity torsion and curvature constraints, 
in contrast to other constructions based on coset techniques. 
This has the advantage that one can consider different 11 dimensional 
backgrounds at once \cite{dWPPS}.  Also we found complete agreement with 
the coset construction in \cite{dWPPS}, where agreement with gauge completion 
results to lowest order in $\theta$ was found. However, the covariant torsion 
and curvature constraints can already be derived from the first order in 
$\theta$ gauge completion \cite{BH} and therefore as this paper only relies 
on these constraints, there is full agreement to all orders in $\theta$. 
The two approaches are therefore completely equivalent. 
In this paper, we restricted ourselves to backgrounds with 
vanishing gravitino and covariantly constant field strengths, but a study 
of more general backgrounds along the same lines would be interesting. 
\par 
Following \cite{RKILL} we considered the gauge fixing of probe M-brane actions 
in $adS_4 \times S^7$ and $adS_7 \times S^7$. Two $\kappa$-gauges
were proposed related to the killing spinors of the background and the
simplification of the geometric superfields was discussed for both cases.
\par 
By construction the gauge-fixed actions will be invariant under 
superconformal transformations because $OSp(4|8)$ transformations, which 
are the isometries of the $adS_4\times S^7$ M2 near-horizon background, and 
$OSp(6,2|4)$ transformations, which are the isometries of the $adS_7\times 
S^4$ M5 near-horizon background, become upon gauge fixing non-linearly 
realized superconformal transformations on the remaining world-volume 
fields.  For the M2 we have a superconformal scalar multiplet in 3 
dimensions and for M5 a superconformal $(0,2)$ tensor multiplet in 6 
dimensions.  The precise form of the transformation rules, which have 
been written down for the (bosonic) conformally invariant actions in 
\cite{CKKTVP}, still have to be derived. 
 
\vskip1cm 
\noindent 
{\sc Acknowledgements}\\ 
It is a pleasure to thank R.~Kallosh, D.~Sorokin and A.~Van~Proeyen for 
very stimulating discussions about various properties of $p$-brane actions, 
worked out in detail for the M branes in this paper. \\ 
This work is supported by the European Commission TMR programme 
ERBFMRX-CT96-0045. 
 
\end{document}